\documentstyle[preprint,aps]{revtex}
\begin{document}
\draft
\title{
{\bf Bound states and transmission antiresonances in
parabolically confined cross structures:
influence of weak magnetic fields}
}
\author{R. Akis$^1$, P. Vasilopoulos$^2$, and P. Debray$^3$}
\address{
1 Center for Solid State Electronics Research, Center for Systems
Science and Engineering, and Department of Electrical Engineering,
Arizona State University Tempe AZ\,\, 85287}
\address{
2 Concordia University, Department of Physics, 1455 de
Maisonneuve Blvd O, Montr\'{e}al, Qu\'{e}bec, Canada, H3G 1M8 }
\address{
3  Service de Physique de l'Etat Condense and Centre National del la
 Recherche Scientific,
 Centre d'Etudes de Saclay, 91191
Gif-sur-Yvette, France}
\maketitle

\begin{abstract}
  The ballistic conductance through a device
  consisting of quantum wires, to which two stubs are attached laterally,
is calculated assuming  parabolic 
confining potentials of frequencies $\omega_w$ for
the wires and $\omega_s$ for the stubs. As a function of the ratio
$\omega_w/\omega_s$ the conductance shows nearly periodic minima
associated with quasibound states forming in the
stubbed region. Applying a magnetic field $B$ normal
to the plane of the device changes the symmetry of
the wavefunctions with respect to the center of the
wires and leads to {\it new} quasibound states in the
stubs. The presence of the magnetic field can
also lead to a second kind of state, trapped mainly in the wires by the 
corners of the confining potentials, that yields
conductance minima as well. In either case, these bound states form for 
{\it weak} $B$ and strong confining frequencies and thus 
are {\it not} edge states. Finally, we show experimental evidence
for the presence of these quasi-bound states.

\end{abstract}

\section{INTRODUCTION}

Technological advances in microfabrication techniques now
allow the manufacture of semiconductor structures that have
dimensions smaller than the
elastic and inelastic mean scattering lengths. In such
{\it mesoscopic} structures,
the electronic transport is {\it ballistic} \cite{1},
and the conductance is governed by the fact that
the electrons behave like quantum mechanical waves.
This is particularly true at low temperatures.

The wavelike behavior of electrons in such structures has led
to the study of devices that are analogous
to those used in microwave technology.
The simplest such device is the quantum wire (QW), which can be thought
of as an electronic waveguide.  Some more complicated structures
involve having
QWs cross to form junctions, attaching finite branches to the QWs so
that they become corrugated, and
connecting the QWs to the electronic equivalent of a resonant cavity.
These closely related structures have generated  experimental
and theoretical interest  \cite{2}-\cite{9}. In particular, resonant tunneling
and quasibound states in
stub and cross structures have been focused on
because in these systems the electrons are
not bound classically by any potential barriers.
In addition, they are unusual in that the presence of quasibound
states can lead to resonant {\it reflection} instead of transmission,
i.e., transmission {\it antiresonances}.

A special type of structure in this class
is the electron stub tuner (EST), in which the length of the stub, laterally  
attached to the QW, would be controlled by an independent gate. 
If the width of the QW and the stub are such that
both allow only a single propagating mode for a given incident energy, then
 the conductance $G$ is a periodic function of the
stub length $c$, with $G$ oscillating between 0 and 1, in
units of ${2 e^2 /h}$, making it potentially useful as a type of
transistor \cite{10}. The conductance minima that result can be attributed
to {\it destructive} interference between the electron waves in the
wire and those reflected from the stub.
A more sophisticated device is the double
electron stub tuner (DEST) depicted in Fig. 1 (a).
If the length of the DEST is
kept {\it fixed} while it is being made asymmetric by suitably
synchronized gate voltages, a conductance output, nearly
square wave in form, can be achieved as a function of  the degree of asymmetry
with potential uses in analog-to-digital
converters \cite{11}.

In this paper we consider a few important issues with
regard to these devices.
First of all, in previous theoretical work on cavity and stub structures,
the confining potentials were always assumed to be infinite square
well in nature \cite{2}-\cite{12}.
However, it is well known that for very
narrow QWs a parabolic potential is
more appropriate \cite{13}. In such a case the
width of the electronic wavefunctions and thus the device dimensions
{\it are not well defined}. After briefly presenting the formalism
in Sec. II, we will present {\it the zero field results}
for a DEST,
which show a nearly periodic conductance as a function of the
ratio $\omega_w/\omega_s$ even under these circumstances. Some of these results
will contrasted with those obtained assuming a square confinement.

Secondly, \lq\lq true" ESTs and DESTs, with independent
gates controlling the stub lengths, have yet to be fabricated.
Thus far, experiments have only been done on cavity structures in which the conductance
was studied as a function of a gate voltage $V_g$ that affected several of
the device dimensions simultaneously  making it difficult to
interpret the results definitively as resulting from the interference
effects mentioned above.  However, $G$ has been measured in
these structures as a function of a {\it weak} perpendicular
magnetic field $B$ as well for fixed
$V_g$ \cite{14}. The results show minima in $G$
as a function of $B$. The values of $B$ and the confining
frequencies are such that edge states do
not occur and so much of the previous work on quantum
dots is not applicable. As we shall show in section III, such minima
can arise in DESTs when electrons are reflected resonantly  from quasi-bound
states in the stubs. We further show that
{\it new} quasibound states are created when $B\neq 0$ that are not present in
zero field. This happens because $B$ changes
the symmetry of the wavefunctions at zero
field with respect to the transverse direction
$y$ and leads to {\it new} couplings between the wire and stub wavefunctions.
In Sec. IV, we show results for actual experiments and interpret them
qualitatively  in terms of those of Sec. III. Conclusions follow 
in Sec. V.

\section{FORMULATION OF THE TRANSMISSION PROBLEM}

In this paper we consider parabolic confinements.
For a parabolic one along the $y$ axis in
the wire ($w$) and stub ($s$) regions, i.e., we take
\begin{equation}
V_{w,s}(y)=m^* \omega^2_{w,s} y^2/2,
\label{1}
\end{equation}
 and/or the presence of an  applied
magnetic field, {\bf B}= (0,0,B). The parabolic confinement in a DEST is depicted
three-dimensionally in Fig. 1 (b).  The narrower parabolas,
defined by the frequency $\omega_w$, represent the two parts
of the QW; the wider parabola, defined by $\omega_s <\omega_w$,
represents the stubbed region. 
The confinement in the stubs along $x$
is achieved essentially through
the difference in stub and wire potentials,
$\Delta V(y) = V_w(y)-V_s(y)$, so that only a {\it finite}
potential barrier
is created. As a result, the electronic wavefunction in the stub
regions will not go to zero at the boundaries
and thus it can
spill over into the QW. 
This
is a feature our present model
shares with saddle potentials that is absent in
infinite square-well models used in past calculations.

To evaluate the transmission through the device depicted in Fig. 1, we
solve Schrodinger's equation on a mesh, using an iterative matrix method. 
We summarize the essentials of the method below and refer to \cite{15}
 for all the mathematical details.
The general situation is one in which the QWs, which are
 connected to the stub structure, extend outward to $\pm \inf$ along the
$y$ direction.
The problem is solved on a 
square lattice of constant $a$. Along the $x$ direction, the
system must be cutoff after a finite number of lattice sites, say 
$M$. Thus, the situation is one in which the parabolic QWs that are
depicted are in fact enclosed within a larger waveguide that is bounded
by infinite potential barriers.
The region of interest, containing the actual stub structure, can be broken 
down into a series
of slices along the $y$ direction. The discrete form of the Hamiltonian relates
quamtum mechanical amplitudes between adjacent slices. Keeping only terms
up to first order in the derivative, this has the form:

\begin{equation}
(E_F-{\bf H}_j)\psi_j+{\bf H}_{j,j+1}\psi_{j-1} +{\bf H}_{j,j-1}\psi_{j+1}=0,
\label{2}
\end{equation}
where $\psi_j$ is a $M$-dimensional column vector 
containing the amplitudes of the
$j$th strip. The matrices ${\bf H}_j$ 
represent the Hamiltonians for the individual
slices. 

By approximating the derivative, the kinetic energy terms of the Hamiltonian 
get mapped onto a tight-binding model with  $t=-\hbar^2/2m^*a^2$ representing 
the nearest-neighbor hopping. To include the effects of the confining potential, 
one adds to the on-site energies,  which occur along the diagonals
of ${\bf H}_j$, the terms $v_{j,m}$, which represent the 
potential on site $(j,m)$ in units of $t$. Parabolic confinement can be modelled 
easily using  Eq. 1 for $v(l,m)$, with 
$y= a(m-(M+1/2))$, so that $y=0$ occurs at the center of each slice. 
The matrices ${\bf H}_{j,j+1}$ and ${\bf H}_{j,j-1}$ give the 
inter-strip
coupling and are related by 
${\bf H}_{j,j+1}={\bf H}_{j,j-1}^*$. We use the gauge in 
which these {\it diagonal} matrices are given by
${\bf H}_{j,j+1}(l,l)=-t\ exp(2\pi i\beta l)$ 
where $\beta=Ba^2/\phi_0$ is the
magnetic flux per unit cell and $\phi_0=\hbar/e$. 
In this gauge, the magnetic field points along the 
$ z$ direction. Equation (2) can be used to derive a transfer matrix
which allows us to translate across the system and thus calculate the 
transmission coefficients which enter the Landauer-B\"uttiker formula to
give the conductance. Transfer matrices however are notoriously unstable
due to exponentially growing and decaying evanescent modes. This problem
was overcome in Ref. 15 by performing some clever matrix manipulations 
and turning
the process of translating across the system into an 
iterative procedure, rather
than multiplying transfer matrices together.
It has been found that the method gives results equivalent to
those of the recursive Green's function technique \cite{15}, which is the
most common approach to this type of problem. Its 
advantage over the latter is that it is conceptually simpler 
and easier to implement. 
Once the procedure is complete, one obtains the transmission
coefficients,
$t_{nm}$,  and reflection coefficients, $r_{nm}$, for the individual modes.
Given these, the amplitudes of the wavefunctions at specific
values of $x$ and $y$ can be obtained by a progressive backward substitution.

The total transmission $T$ is  given by
\begin{equation}
T= \sum_{nm} T_{nm} = \sum_{nm} |t_{nm}|^2  {v_n \over v_m},
\label{9}
\end{equation}
where $v_n$ and $v_m$ correspond to the velocities of the
transmitted and incident modes respectively and 
the sum is over {\it propagating} channels only. The
conductance $G$ then at zero temperature
is given by the Landauer-B\"uttiker formula:
$G=(2e^2/h)T$.

As one reduces the lattice spacing in the
discrete model, so one is in the limit where
$a \ll  \lambda_F$, $\lambda_F$ being the
Fermi wavelength, the results
that are obtained eventually can be mapped to those
of the continuous case.
Since we work in the regime of
small $a$, it will 
be convenient to make reference to the modes that occur in the continuous case.  
The nth channel wavenumber $\alpha_n$ in the wire is
\begin{equation}
\alpha_{n}={\Omega_w \over \omega_w} \sqrt{
{2m^* \over \hbar^2} \left [ E_F-(n+1/2)\hbar \Omega_w \right ]}
\label{11}
\end{equation}
Similarly, in the stub region, the wavenumber $\gamma_m$ takes the
form:
\begin{equation}
\gamma_{m}={\Omega_s \over \omega_s} \sqrt{
{2m^* \over \hbar^2} \left [ E_F-(m+1/2)\hbar \Omega_s \right ]}
\label{12}
\end{equation}
Here $\Omega_{s,w}^2=\omega_{s,w}^2+\omega_c^2$ and $\omega_c=|e|B/m^*$ is
the cyclotron frequency.
The {\it modes} $\phi_n$ along the y-axis
depend on whether the waves are traveling
in the positive ($exp(i \alpha_n x)$) or
negative ($e^{-i \alpha_n x}$) x-direction. We thus have wire modes,
$\phi^{w\pm}_n(y)=\varphi^{w}_{n}(y \mp
(\hbar \omega_c/ m^* \Omega^2_w)\alpha_n)$,
and stub modes,
$\phi^{s\pm}_m(y)=\varphi^{s}_{m}(y \mp
(\hbar \omega_c/ m^* \Omega^2_s) \gamma_m)$,
where $\varphi^\omega_j(y)$ is the jth
harmonic oscillator (HO) wave function. Notice that for $B=0$, we have
$\varphi^+(y) =\varphi^-(y)$.

\section{RESULTS}
\subsection{Zero field}

	In previous theoretical work on ESTs, with stub length 
$c$ and width $b$, a periodic conductance output has been 
obtained, as a function of $c$,
for infinite square-well confinement; the period $\delta c$ 
is given by
\begin{equation}
\delta c=  \pi / \sqrt{ 2 m^\ast E_F/\hbar^2 -
\biggl(\pi/ b \biggr)^2}=\lambda_s/2
\label{13}
\end{equation}
when only one mode is allowed in the QW and stub regions.
Equation (13) is a restatement of the condition for {\it destructive}
interference, $k_s \delta c = \pi$, since $\lambda_s=2\pi/k_s$ is the
electronic wavelength along  the stub. Notice that the period
{\it increases}
as $b$  is made smaller. For a {\it symmetric} DEST \cite{16},
this period is doubled, so that $\delta c = \lambda_s$.

An interesting question is whether or not the conductance remains
periodic if the confinement is instead {\it parabolic}, particularly
when considering that in this case  the stub length is no longer
well defined. In the pertinent literature
it is quite common to use the classical turning points to define an
{\it effective} halfwidth $W_{eff}$ of
the parabolic well through $E_F=m^* \omega^2W^2_{eff}/2$. Taking
$\omega=\omega_s$ gives an effective stub length
 \begin{equation}
c_{eff}= 2 W_{eff} = 2 \sqrt{2 E_F \over m^*} {1 \over \omega_s}.
\label{14}
\end{equation}
If the DEST in the parabolic case behaves in a
manner similar to that of past calculations, one might expect then
that the conductance $G$ of a DEST to be
a periodic
function of $1/ \omega_s$ for
fixed $E_F$.
As we show in Fig. 2 (a), this
is in fact the case. We plot $G$  as a function
of $\omega_w/\omega_s$
for {\it fixed}  $\hbar \omega_w=6.39$ meV
and
$E_F=9$ meV so that there is one propagating mode in the
connecting quantum wires. The width of the stub is $b= 400 \AA$ (solid curve)
and $b= 350 \AA$ (dashed curve). When
$b$ decreases  the period increases; this is consistent with the 
results for infinite square-well confinement as expressed in Eq. ($\ref{14}$).

The transmission minima displayed in Fig. 2 (a) can be considered
to occur as a result of destructive interference. An alternate
but complementary point of view is that they occur as a result
of resonant {\it reflection} from quasi-bound states in the
stubbed cavity. They are transmission {\it antiresonances}. This is
illustrated in Fig. 2 (b), where $|\psi(x,y)|$ is plotted as a function
of $x$ and $y$. To generate this plot, we have set $E_F=9$
meV, $b= 400 \AA$ and set
$\hbar \omega_s=2.89$ meV so that
$\omega_w/\omega_s= 2.21$. The picture
corresponds to the {\it first} transmission minimum in the solid curve
in Fig. 2 (a).  A standing wave corresponding to a quasibound state
is apparent in the cavity  region between the arrows along the x axis.

Further insight into the antiresonances is obtained as follows.
Since only one mode is occupied in the quantum wires,
the full wavefunction $\phi(x,y)$ goes as the $n=0$ HO
wavefunction, $\varphi^{\omega_w}_0(y)$ for a set value
of $x$. What is interesting
is that the standing wave in the cavity
region, despite being obtained by summing
over the contributions of many HO wavefunctions,
can be associated with the $n=2$ HO
wavefunction $\varphi^{\omega_s}_2(y)$. In particular, if
we set $x=x_\circ$, where $x_\circ$ represents {\it the center
of the stub} ($200 \AA$ in this case),
then $\psi(x_\circ,y)$ can be fit almost perfectly
by using $\varphi^{\omega_s}_2(y)$ {\it alone}. While this is
not true away from $x=x_\circ$, $\psi(x,y)$ in the stub region
keeps the basic $n=2$ HO form  and thus it
remains {\it even} with respect to $y=0$,
the center of the quantum wires.
Consequently, the conductance minima or antiresonances
can be attributed to an {\it even-even} coupling between the
$n=0$ state in the wire and the $n=2$ state bound in the
stubbed cavity or DEST.
The other minima in the solid curve of Fig. 2 can similarly
be associated with an even-even coupling between the 
$n=0$ and $n=4, 6$, etc. states. Coupling between the
even, in the wire, and odd, in the stub,
HO states does not occur because they are orthogonal to each other.

So far the results are similar to those obtained for a
square-well confinement. The main difference between them is
that the evanescent modes {\it in the connecting QW's} in the square-well 
case decay very slowly. Thus, a long exponential \lq\lq tail" is left in the
wave-function in the exiting QW, even if there is 100 percent
reflection of the incident propagating mode. 
This would be a major liability in the fabrication of an operating device, 
since  the presence of the \lq\lq tail" may result in resonant tunneling  rather
than resonant reflection thus making it difficult to produce a device that
actually produces the desired effect. 
No such tail is apparent in the figure. The fast decay of the 
evanescent modes in the case of a parabolic potential
is related to the wavenumbers given by Eqs. (4) 
and (5) rather than by Eq. (6).

\subsection{Finite field}

{\it 1. Offset or field}

We now consider  a finite but {\it weak} magnetic
field $B$. By {\it weak} we mean a field that is not strong enough
to push the wavefunctions completely
over to one side. We are {\it not}
in the edge state regime. The use of the term
\lq\lq weak" is appropriate to the experimental situation
described in Sec. IV, where the dimensions of the experimental
samples were several hundred to a few thousand $\AA$, which is our motivation.
For a QW with a $c_{eff}$ of a few
hundred $\AA$, one expects $\omega_c \ll \omega_w$
for $B < 1 T$. In addition, this regime has
been much less explored than the edge-state regime.
For simplicity we neglect the Zeeman splitting.

   In Fig. 3 (a), we again plot $G$ as a function $\omega_w/\omega_s$ for
fixed $E=9$ meV, $\hbar \omega_w=6.39$ meV and
$b= 400 \AA$, for three different situations, the upper two curves
offset by $G=1$ and $G=2$, respectively, for clarity.
The bottom curve is the same as the solid curve in Fig. 2 (a).
For the middle curve, we have put in a small offset, $d= 20 \AA$,
so that the DEST is now {\it asymmetric}, with potential
$V_{DEST}(y) \rightarrow  m^* \omega_s^2 (y+d)^2/2$.
 We see that with the
asymmetry  the antiresonances that occur in the symmetric
case are now shifted down slightly. Secondly, and more importantly,
a whole new set of antiresonances occur in between the original
minima. These occur due to the the breaking of symmetry of the wave functions, 
allowing the even $n=0$ QW state to now couple with the odd states
($n=1,3,5\cdot\cdot\cdot$) trapped in the DEST. Very similar
behavior has been noted in the case of square-well confinement.
The upper curve is for a {\it symmetric} DEST, but now in the presence
of a finite magnetic field, $B=0.3$ T. We see that the presence
of the magnetic field produces much the same result as the asymmetry-
the shifting of the original antiresonances, and the appearance of
the new set of minima at virtually the same locations. In Fig. 3 (b)
we  plot $|\psi(x,y)|$ for $\omega_w/\omega_s=4.935$ and $d= 20 \AA$
($G \sim 0$ for these parameters).
Here, the antiresonance wave-function has six lobes, indicating the
coupling of a $n=5$ odd state in the DEST with the $n=0$ even state
in the QW in this case. The corresponding wave function in the presence
of a magnetic field is shown in Fig. 3 (c) for 
$\omega_w/\omega_s=4.896$ and $B= 0.3 T$. The state shown in
this picture is almost indistinguishable from the previous one. Interestingly,
the most significant difference between the two pictures occurs in
the incident waves.  In the finite field case
a standing wave appears that is quite similar to the one evident in
Fig. 2 (b). In the asymmetric case, the waves have a more irregular
appearance. One obtains similar
results for the other even-odd antiresonances. 

Given these results, we conclude the coupling 
between even and odd states in the presence of a magnetic field
occurs here because, when $B$ is finite, {\it the symmetry
about $y=0$ is broken}. Noting that the wavefunction in
Fig. 3 (c) appears almost completely symmetric about $y=0$, it is
obvious that the presence of edge states is not required
for this coupling to take place. In fact, it can occur
for arbitrarily small $B$. However, the smaller $B$ is, the 
narrower the even-odd antiresonances that occur in Fig. 3(a)
become. Another important point is that the position of the
antiresonances depends on the value of the magnetic field.
For example, the first antiresonance, which corresponds to 
a $n=0$ QW-$n=1$ DEST coupling, occurs at $\omega_w/\omega_s=1.1164$
for $B=0.11 T$, $\omega_w/\omega_s=1.17$ for $B=0.29 T$ and
$\omega_w/\omega_s=1.18$ for $B=0.46 T$.
This shifting of the resonance as a function of $B$ for
different choices of $\omega_w/\omega_s$ can be understood,
at least in part, in terms of the lining up of the energy
level of the bound state of the cavity, $E_{bound}$,
with that of the incident
electrons, $E_F$, which is necessary for
a resonance effect to occur. From our previous discussion about fitting
the wavefunction in the DEST, it is apparent that the energy level
structure of the quasibound states is tied to $\Omega_s$.
A larger (smaller) value of $\omega_w/\omega_s$ means that $\omega_s$
is smaller (larger), thus a larger (smaller) value of $B$ is required to
ensure that $\Omega_s$ remains at the  value that lines up the
Fermi level with the bound state
level. This argument, however, is somewhat oversimplified
in that the bound state energy is not determined by
$\Omega_s$ alone. The bound states are confined along
{\it both} the
$x$ and $y$ directions and so the $x$ confinement must neccessary
contribute to the energy of the $n=1$ bound state,
so that we should have
$E_{bound}= 3 \hbar \Omega_s/2 + E_x$. However, as the confinement
along $x$ is incomplete and the system is open, the contribution
$E_x$ is
difficult to quantify, at least analytically.
Importantly, as $B$ changes $\Omega_s$, the
confining potential in the stub along $x$ is
also being altered, thus complicating the physical picture.
As a result, the value of $\Omega_s$ for which antiresonance
occurs is slightly
different for different values of $B$.
The lining up of QW and DEST energy levels is also the likely
explanation of the observed downward shift in both the finite
$B$ and finite $d$ cases.

{\it 2. Offset and field}

In Fig. 4 (a) we plot $G$ vs  $B$ for fixed $\omega_w/\omega_s=4.9$ 
that corresponds to the $n=0$-$n=5$ antiresonance.
The solid curve corresponds to the DEST being symmetric.
The broad minimum at about $\sim 0.28$ corresponds to the
antiresonance in question. It is interesting to see
what happens when {\it a finite $B$ and a finite offset} are
present at the same time, as individually they appear
to have similar effects. The dashed and dotted curves
correspond to $d= 20 \AA$ and $d= 40 \AA$, respectively.
Oddly, the conductance minima become shallower for
increased $d$, as if the magnetic field and asymmetry 
are canceling each other out. Importantly, essentially
the same curves are generated if we replace $d$ with
$-d$. A clue to this behavior can be seen in Fig. 3 (a).
While the antiresonances  occur at essentially the
same spots, the lineshapes are different, with
$G = 1$  followed by $G= 0$ in the case of
finite $d$, and almost exactly the {\it mirror opposite} 
for finite $B$.
In either case, the lineshapes are asymmetric, that is, they 
are of {\it Fano} type. The occurence of Fano antiresonances in stub
structures has been the subject of several papers, typically using
simple qualitative models \cite{5}-\cite{7} (stub and wire both treated
as being purely one dimensional). Stub structures, unlike say
a double barrier problem, yield {\it both} transmission {\it poles}
in the complex energy plane, the real part of 
which is associated with the energy of the quasibound
states and yield unit transmission, and transmission 
{\it zeroes} (the antiresonances). If the pole and the zero do 
not occur at the same location in energy, one obtains the 
asymmetric Fano lineshape. This gives a $G=1$ 
peak followed by a $G=0$ minimum when $E_{pole}<E_{zero}$, and visa
versa when $E_{pole}>E_{zero}$. Figure 3 (a), however, shows the 
antiresonances as a function of $\omega_w/\omega_s$, which
we remind the reader is a measure of stub length for fixed $\omega_w$. 

The \lq\lq flipping" of the Fano shaped antiresonance also 
occurs with respect to {\it energy} and this is shown
in Fig. 4(b), where $G$ vs. $E$ is plotted for
fixed $\omega_w/\omega_s=4.9$. Once again, the
minima here correspond to the $n=0$ - $n=5$ antiresonance. The
solid curve corresponds to $B=0.28 $ T and $d=0$, while the
dashed curve is for $B=0$ and $d=20 \AA$. Note that the
conductance minima occur at slightly different locations.
The dotted curve has both $B=0.28 $T and 
$d= 20 \AA$, which shows the hybrid lineshape, the 
result of the \lq\lq competition" between the two sources 
of symmetry breaking. In the region of the
minimum, this third curve looks somewhat like 
an average between the other two curves. We note
that the conductance maximum follows the minimum
in the combined curve, like the finite $B$ only curve.
We note that the minimum is much wider for the finite $B$
only curve than for the $d$ only curve, indicating that
the finite $B$ is producing a stronger effect in comparison 
to the finite $d$ in this case, and is essentially winning out.
Again referring back to Fig. 3 (a), we note that the 
\lq\lq flipping" effect does not occur when the  
field $B$ is turned on for the even-even antiresonances, presumably
because we consider a relatively weak field $B$. 

{\it 3. Two conductance minima}

In Fig. 5 (a) we again plot $G$ vs $B$. However, unlike the
previous example, {\it two} transmission minima are apparent 
for each of the curves shown here. The solid, dashed and dotted
curves correspond to 
$\omega_w/\omega_s=3.0, 2.91$, and $2.85$, respectively.
In Fig. 5 (b),  
$|\psi(x,y)|$ is plotted as a function
of $x$ and $y$ for the first minimum in the 
$\omega_w/\omega_s=3.0$ curve, which occurs at $B=0.27$ T.
Unlike the previous wave function plots, we are looking directly
from above  and higher amplitudes are represented
by darker shading. The incident electron waves are traveling
from the top to the bottom in this picture.
The quasibound state in this case has four
lobes along the length of the stub and thus represents coupling
between $n=0$ and $n=3$ states and is yet another example
of the even-odd coupling phenomenon we have already pointed
out. More interesting is
the wave function that corresponds to the second minimum
at $B=0.67 $T, which is plotted in Fig. 5(c). Here 
the wave function again has four lobes, but in this case
there are two each in {\it both} the $x$ and $y$ directions.
The quasibound state shown here does not arise from
confinement by the stubs, but is held in place by the corners
formed by the intersection points of the stub and wire
potentials. Quasibound states of this type were first
found to occur theoretically in intersecting
quantum wires in a situation analogous to having stubs
of infinite length by Schult, Ravenhall and Wyld \cite{3}.
They  pointed out
two such \lq\lq intersection" states, the lower energy state 
consisting of 
one large lobe in the intersection region, occuring below the 
of the first propagating mode of the quantum wires, and a
four-lobed
excited state having the same odd symmetry of the state we see
here.   

In the curve for 
$\omega_w/\omega_s=3.0$, the two minima have a relatively large
spacing in B. When $\omega_w/\omega_s=2.91$, the minima are
quite closer to each other, with the lower minimum occuring
at a higher value of $B$, while the second one
remains fixed. In fact, this is as close to each other as the minima
get and they never merge for any value of
$\omega_w/\omega_s$. This is a situation akin to an
anticrossing from band structure theory.  
The wave functions for these two minima are shown in
figures 5 (d) and (e). These wave functions are virtual
mirror images of each other and appear to be {\it hybrids} of
the stub-confined and intersection-confined states shown
in the previous two panels. 

For $\omega_w/\omega_s=2.85$, the second minimum occurs
at $B=0.8 $ T a somewhat higher value of $B$ than the
previous two cases, while the first minimum occurs at
$B=0.57 $ T. The wave function corresponding to the first minimum
of this curve is shown in Fig. 5 (f). It is virtually a mirror reflection
of the intersection-confined wavefunction shown in
Fig. 5 (c). The wave function for the second minimum in
this case is shown in Fig. 5 (g) and again has the hybrid
form.  

As is evident from our results, the relative positions
of the two minima depend  quite sensitively on 
$\omega_w/\omega_s$. It should be pointed out that,
when $\omega_w/\omega_s$ is increased $3.05$, the lower 
conductance minimum no longer occurs  leaving only the 
intersection-confined state at approximately the same position
as it is for $\omega_w/\omega_s=3.0$. 
On the other hand, if $\omega_w/\omega_s$ is decreased further 
below $2.85$, the
position of the lower minimum, which now corresponds to the
intersection-confined state, occurs at lower and lower values
of $B$, but it shifts less significantly than the second minimum 
which occurs at increasingly higher $B$ values. That the
intersection-confined state is less sensitive to changes in 
$\omega_w/\omega_s$ is not surprising, since its presence
should not depend too strongly on stub length. On the other
hand, the reason why there is a shift at all in its position, when
$\omega_w/\omega_s$ is changed, is because while we are
changing the stub length, we are also changing the confinement
at the corners as well in our model.

\section{EXPERIMENTAL EVIDENCE FOR QUASI-BOUND STATES}

In this section, we present experimental results which lend support to our 
theoretical analysis and provide evidence for the presence of quasibound 
states in a DEST device and the appearance of new transmission minima 
under the influence of a magnetic field applied perpendicular to the 
device plane. Some preliminary results and details of sample fabrication 
and experimental measurement technique have been reported 
earlier\cite{14}. The 
DEST device was fabricated using Schottky gates to define device geometry 
from a high-mobility ($\mu =110 m^{2}/V_{s}$ at 4.2K) and low-electron-density 
($n = 3.1x10^{15} m^{-2}$) AlGaAs/GaAs modulation-doped (Si) heterostructure 
grown by  MBE and is shown in the inset of Fig. 6. The Fermi energy of the 
2DEG was measured to be $E_{F} = 8.50$ meV. The lithographic dimensions 
of the device were : $a = b = 2500 \AA$, 
$c = 8500 \AA$, and $l = 1500 \AA$, respectively, $l$ being 
the length of the connecting wires. Figure 6 shows the conductance $G$ of 
the device in the absence of magnetic field measured as a function of gate 
voltage $V_{g}$ at 70 mK. This temperature is a small fraction of 
$E_{F}$
to be considered essentially zero. As $V_{g}$ is made more negative, the device 
dimensions $a$, $b$, and $c$ all decrease at the same time due to depletion. 
From measurements of the quantized conductance plateaus of a single 
quantum wire with lithographic width the same as that of the DEST wires, 
it was found that at $V_{g} = -500 mV$ the Fermi level lies just below the 
bottom of the second ($n = 1$) wire subband, and the corresponding wire 
width is $400 \AA$, so that for $V_{g}$ 
( $-500$ mV one could say that transport is 
in the fundamental mode of the connecting quantum wires and only the 
lowest ($n = 0$) wire subband is occupied. Assuming the depletion at the 
stub edges is the same as that at the wire edges as the gate voltage is 
decreased, a rough estimate of the DEST dimensions at $V_{g} = -500$ mV could 
be obtained : $a = b = 400 \AA$, 
$c = 6400 \AA$. Though the estimate is rough, we 
can safely expect quite a few DEST subbands to be occupied. Since the 
Fermi level is the same across the device and the Fermi energy does not 
change with $V_{g}$, a decrease in $V_{g}$ accompanied 
by corresponding reduction of 
device dimensions means a decrease in the effective wire width and stub 
length as derived from the definition of classical turning points and 
given by Eq. (7). One could then say that the effective wire and stub 
confining frequencies increase as the gate voltage is made more negative. 
Since the depletion at the gate edges ( $\simeq 2.9 \AA/mV$) is the 
same for the 
wires and the stub, a change in $V_{g}$ over a small range brings about
little 
relative change in the stub length. However, for the wires, because of the 
much shorter dimension, the relative change in the wire width is quite 
important as $V_{g}$ is swept. Considering the $V_{g}$ range 
between $-500 mV$ and 
pinch-off, one could then possibly consider the stub confining frequency 
$\omega_{s}$ to stay practically constant, while the wire 
confining frequency $\omega_{w}$ to 
increase rapidly with decreasing $V_{g}$. In Fig. 6, therefore, decreasing 
$V_{g}$ 
would mean increasing the ratio $\omega_w/\omega_s$. It would also mean
 sweeping the 
Fermi level down across the stub subband levels given by $\omega_{s}$
. In Fig. 6, 
the conductance $G$ shows two prominent minima and three maxima for 
$V_{g}$ less 
than $-500$ mV. The observed minima can be attributed to an even-even 
coupling between the $n = 0$ state in the wire and the $n =$ even quasi-bound 
states in the stubbed cavity or DEST, as the Fermi level sweeps down the 
stub energy level structure. This analysis is in line with the theoretical 
prediction of the previous section and the observed minima can be 
considered as an experimental support of the theoretical analysis 
illustrated in Fig. 2(a). Note that in the present device geometry as 
$V_{g}$ 
changes, the stub width changes as well. The observed minima are thus 
expected to be much broader than the theoretically predicted ones for a 
constant stub width. Moreover, the stub shape may also depend somewhat on 
the gate voltage. The shallowness of the minima can be attributed to 
asymmetry and/or defects\cite{17}, while values of the maxima less than 
$2 e^{2}/h$ 
can be attributed to backscattering at the wire entrance and/or 
impurities. For $V_{g}$ larger than $-500$ mV, transport in 
the wire and in the 
stub since $a \simeq b$ is multimode. The resulting enhanced mixing 
between different modes will result in a more irregular $G$-curve and may 
cause the regular oscillations observed below $V_{g} = -500$ mV to be be 
gradually washed out as seen in Fig. 6.
Based on the above analysis, we could index (n) the minima and maxima of 
Fig. 6. The indexing is shown by arrows. Using the known value of EF and 
the above indexing, we get, for $V_{g} = -500$ mV, 
$h \omega_{w} = 5.67$ meV and $h \omega_{s} = 
1.030$ meV, giving $ \omega_w/\omega_s= 5.50$. This value is close to 
that used to 
generate Fig. 4(a). Note also that at this gate voltage 
$a = b = 400 \AA$. 
Figure 7 shows how the conductance maximum (index 5) of the DEST at 
$V_{g} = -500$ mV,  changes under the influence of a magnetic field applied 
perpendicular to the plane of the device. We have added to Fig. 7, for 
comparison purposes, the theoretical curves of Fig. 4(a) which correspond 
to $b = 400 \AA$. As the field is increased, experimental G decreases and goes 
through a pronounced dip which corresponds to a transmission minimum. The 
minimum in G occurs at $B=0.29$ T, a value that is not strong enough 
to produce edge states. The experimentally observed minimum follows 
remarkably well the B-dependence predicted by theory and may be understood 
in terms of the formation of a new quasi-bound state due to even-odd 
coupling induced by a weak magnetic field as discussed above. The 
shallowness of the observed dip may be due to asymmetry of the 
experimental DEST as illustrated by the theoretical curves THA20 and 
THA40, respectively. The fabrication of a perfectly symmetric DEST is a 
matter of chance and can not be ˆ priori guaranteed. The presence of 
disorder may be playing a role as well.

Support for the transmission antiresonances predicted
in the last section is provided by the  experimental results
of Ref. \cite{14}. The conductance $G$ of a DEST device, as a
function of a perpendicular magnetic field $B$,
shows a deep minimum apparent in the main Fig. 5 of Ref. \cite{14}.
The device was fabricated from high-mobility AlGaAs/GaAs
modulation-doped heterostructures grown by MBE using the
split-gate technique. The gate voltage $V_g$ was so adjusted that 
transport was in the fundamental mode in the quantum wire with the
Fermi level ($E_F=9$ meV) lying just below the second subband
and transmission was unity in the absence of $B$.
The experiments were performed at 70 mK, which
is a small enough fraction of $E_F$ to be considered
essentially zero temperature. The device
dimensions under these conditions were estimated
to be: wire width w = 480 $\AA$ and total DEST 
length c = 6500 $\AA$. This however is a very rough
estimate. At any rate, we expect $\omega_w$  to be considerably larger than 
$\omega_s$ and thus many DEST subbands to be occupied.

The conductance minimum  occurs for $B=0.29\ $T, a value that is 
not strong enough to produce edge states.
The large experimentally observed minimum in $G$ may be
understood in terms of the even-odd coupling and the formation
of bound states discussed above.  The minimum also
shows some superimposed fine structure which may
result from the presence of disorder in the quantum wire
and/or in the DEST, which has been found theoretically
to produce the type of jagged curve shown here \cite{17}.
In addition, we note  the appearance of narrower minima 
superimposed on top of the main one. While these may
be disorder-induced noise, they may indicate
the presence of more than one quasibound state, perhaps combinations of the
intersection-confined and stub-confined states shown in
Fig. 5. It is impossible to say at this point to what kind
of state corresponds the main minimum in this experimental example.
 
The lowest point of the main dip in Fig. 6 shows about 25$\%$ transmission 
whereas there is no transmission ($G \approx 0$) at the corresponding
minimum of the solid curve in Fig. 4. Again this may
by attributable to disorder or the sample's asymmetry
as suggested by Fig. 4.

\section{CONCLUSIONS}

We have calculated the conductance for stubbed electron waveguides
defined by a parabolic potential. In the absence of a  magnetic field we find 
a {\it periodic} conductance output as the stubbed cavity is made longer, which 
is consistent with previous theoretical work done assuming infinite square well
potentials. The conductance minima  or {\it antiresonances}
correspond to quasibound states in the stubbed regions. When the two 
parabolas representing the wire and stub confining potentials are 
displaced with respect to each othe, the symmetry of the wave 
functions, with respect to the center of the wire, is broken and
{\it new} quasibound states occur in the intersection regions.  
The same holds when the two parabolas are not
displaced but a {\it weak} magnetic field $B$ is present because 
the field too  breaks this symmetry thus allowing states in the cavity and
wire, that were previously orthogonal, to couple.
The appearance of these quasibound state is heralded by one or more
dips in the conductance as a function of magnetic field. We emphasize 
that these dips occur in short and long stubs, i.e, whether there are just 
a few or  many stub subbands occupied for electrons incident at the 
Fermi energy. Such dips have been observed experimentally
in electron waveguides with stubbed cavities$^{14}$.

We   have also investigated more sophisticated models for the confinement
potentials, in particular models in which the transition
between the quantum wire and stub regions is made gradually
instead of abruptly as well as combinations of flat
and parabolic confinement. We find that for the most part the
 results are qualitatively similar to those
of the simple double parabolic model shown here. Importantly,
most quasibound states, that occur when the transition
in confinement is not abrupt, tend to be variations of
the hybrid type discussed in the context of Fig. 5.
In addition, we find that it is much more difficult to
get the conductance minima at the low values of B considered
here when all potentials are defined by infinite square-well 
confinement. Unless there is some rounding of the
potentials, as one expects in real devices, the energy
level spacing is too large to permit it.  

\acknowledgements

The work of R A and P V was supported by NSERC grant No. OGP0121756.

\newpage

\begin{figure} 
\caption{ (a) A stubbed cavity of  width $b$ connected to two quantum wires.
(b)  The confining potential in the wires and the
stubbed cavity. The  picture is generated with
$\hbar \omega_w = 6.39$ meV, $\hbar \omega_s= 2.8$ meV,
and $b= 300 \ \AA$. The x range is from $-300 \ \AA$ to $600 \ \AA$
and the y range from $-800 \ \AA$ to $800 \ \AA$.
\label{fig1}}
\end{figure} 

\begin{figure} 
\caption{ (a) Conductance $G$ vs $\omega_w/\omega_s$ for $b=400 \ \AA$
(solid line) and $b=350 \ \AA$ (dashed) with {\it fixed}
$\omega_w=6.39$ meV and $E=9$ meV.
b) A three-dimensional plot of $|\psi(x,y)|$ vs $x$ and $y$ for
$b=400 \ \AA$
and $\omega_s=2.88$ meV. This corresponds to the first minimum in
the solid curve in (a). The two arrows on the bottom right
indicate the edges of the cavity and those on the
left  the width $W_{eff}$ of the quantum wire.
\label{fig2}}
\end{figure} 

\begin{figure} 
\caption{ (a) Conductance $G$ vs
$\omega_w/\omega_s$ for $b=400 \ \AA$,
$\omega_w=6.39$ meV, and $E=9$ meV. The bottom curve is
for a symmetric DEST at $B=0$ T. For the middle curve, offset by $G=1$,
the DEST has been made {\it asymmetric} by a factor of $d= 20 $.
For the top curve, offset by $G=2$,  a $B=0.3$ T has been applied.
Notice the additional
antiresonances that occur in the presence of finite asymmetry and magnetic field.
(b)$|\psi(x,y)|$ vs $x$ and $y$ is plotted for $\omega_w/\omega_s=4.935$
and $d= 20 $. This quasibound state corresponds to the $5th$ minimum in 
the middle curve in (a).
 (c) As in (b) but for $B=0.3$ T and $\omega_w/\omega_s=4.896$.
 This state corresponds to the $5th$ minimum in  the top curve in (a).
 \label{fig3}}
\end{figure} 

\begin{figure} 
\caption{ (a) Conductance $G$ vs $B$ for
$b=400 \AA$,$\omega_w/\omega_s=4.9$.
The solid curve is for a {\it symmetric} DEST and the dashed and dotted 
curves for an {\it asymmetric} one with $d= 20 $ and  $d= 40 $, respectively.
(b) Conductance $G$ vs $E$ for
$b=400  \AA$ and $\omega_w/\omega_s=4.9$.
The solid curve is for $B=0.28$ T and $d=0$ and the dashed one for
$B=0$ and  $d= 20 $. The dotted curve is for $B=0.28$  T and  $d= 20 $. 
\label{fig4}}
\end{figure} 

\begin{figure} 
\caption{(a) Conductance $G$ vs
$B$. The solid, dashed, and dotted
curves correspond to 
$\omega_w/\omega_s=3.0, 2.91$, and $2.85$, respectively.
Note that two conductance minima occur in each  curve.
(b) In panels (b)  through  (g) the wave functions corresponding to these
minima are plotted vs $x$ and $y$ with darker shading corresponding
to higher amplitude. Panels (b) and (c) correspond to the first
and second minima, respectively, for $\omega_w/\omega_s=3.0$; (d) and (e)
correspond to $\omega_w/\omega_s=2.91$  and (f) and (g) to
$\omega_w/\omega_s=2.85$.
\label{fig5}}
\end{figure} 

\begin{figure} 
\caption{ 
Conductance $G$ as function of gate voltage $V_{g}$
 for a nominally symmetric DEST at 70 mK. The numbers accompanied 
 by arrows give stub subband indices. The inset shows a schematic 
 drawing of the DEST geometry as defined by lithography. The hatched areas 
 (G) represent Schottky gates.
\label{fig6}}
\end{figure} 
 
\begin{figure} 
\caption{ 
Conductance $G$ as function of 
magnetic field $B$ applied perpendicular to device 
plane for the DEST shown in Fig.6 at fixed $V_{g}= -500$ mV and 70 mK. 
The theoretical curves THS, THA20, and THA40 are reproduced from 
Fig. 4(a). THS : symmetric DEST, THA20 : with offset 
20 \AA, THA40 : with offset 40 \AA . See text for details.
\label{fig7}}
\end{figure} 


\begin{references}
\bibitem{1} For review articles on the subject  see
C. W. J. Beenaker and H. van Houten,
Solid State Phys. {\bf 44}, 1 (1991), and S.E. Ulloa, A. MacKinnon,
E. Casta\~no, and G. Kirczenow in {\it Handbook on Semiconductors},
Vol. 1, ed. P.T. Landsberg (Elsevier, Amsterdam, 1992).

\bibitem{2} F. M. Peeters, Superlatt. Microstruct.  {\bf 6}, 217 (1989).

\bibitem{3} R. L. Schult,  D. G. Ravenhall, and H. W. Wyld,
Phys. Rev. B {\bf 41}, 12760 (1990). 

\bibitem{4} J. J. Palacios
and C. Tejedor, Phys. Rev. B  {\bf 48}, 5386 (1993). 

\bibitem{5} W. Porod, Z. Shao, and C.S. Lent,
Appl. Phys. Lett.  {\bf 61}, 1350 (1992).

\bibitem{6} P. J. Price,  Appl. Phys. Lett. 62, 289 (1993).

\bibitem{7}  E. Tekman and P. F. Bagwell,
Phys. Rev. B  {\bf 48}, 2553 (1993).

\bibitem{8} Z-L. Ji and K-F. Berggren, Phys. Rev. B
{\bf 43}, 4760  (1991). 

\bibitem{9} M. Leng and C.S. Lent,
Phys. Rev. Lett. {\bf 71}, 137 (1993).


\bibitem{10}  A. B. Fowler, U.S. Patent No. 4, 550,330 (Oct. 29,1985);
F. Sols, M.Macucci, U. Ravaioli and
K. Hess, Appl. Phys. Lett. {\bf  54}, 350,
(1989); S. Datta, Superlatt. Microstruct. {\bf 6}, 83 (1989).

\bibitem{11} P. Debray, R. Akis,  P. Vasilopoulos, and J. Blanchet,
Appl. Phys. Lett.  {\bf 66}, 3137 (1995).

\bibitem{12} H. Wu, D. W. L. Sprung, J. Martorel,l and
S. Klarsfeld,  Phys. Rev. B {\bf 44}, 6351  (1991).

\bibitem{13} A. Kumar, S. E. Laux, and F. Stern,
Appl. Phys. Lett.  {\bf 54}, 1270 (1989).

\bibitem{14} P. Debray, J. Blanchet, R. Akis, P. Vasilopoulos,
and J. Nagle, Inst. Phys. Conf. Ser. {\bf 141} 835 (1995).

\bibitem{15} T. Usuki, M. Saito, M. Takatsu, R. A. Kiehl, and N. Yokoyama, 
 Phys. Rev. B {\bf 52}, 8244 (1995).

\bibitem{16} R. Akis,  P. Vasilopoulos, and P. Debray,
Phys. Rev. B {\bf 52}, 2805  (1995).

\bibitem{17} H. Sordan  and K. Nikolic,
Phys. Rev. B {\bf 52}, 9007  (1995).
\end{references}
\end{document}